\documentclass[%
aip,
rsi,%
amsmath,amssymb,
reprint,%
groupedaddress
]{revtex4-1}

\usepackage[pdftex]{graphicx}
\usepackage[eulergreek]{sansmath}
\usepackage{color}
\usepackage{braket} 
\usepackage{tikz}

\definecolor{darkred}{rgb}{0.6,0.,0.}
\definecolor{darkgreen}{rgb}{0.,0.5,0.}
\definecolor{darkblue}{rgb}{0.,0.,0.6}
\usepackage[ breaklinks, colorlinks=true, linkcolor=darkred, citecolor=darkgreen, urlcolor=darkblue]{hyperref}

\usepackage[inactive,tightpage]{preview}
\PreviewMacro[*[[!]{\inputgnuplottex}
\PreviewMacro[*[[!]{\includegraphics}

 \newcommand{\diffd}{\text{d}}
\newcommand{\dr}{\mathrm{d}\mathbf{r}}
\renewcommand{\vec}[1]{\mathbf{#1}}
\newcommand{\punc}[1]{\,#1}

\newcommand{\secref}[1]{Sec.\,\ref{#1}}

\graphicspath{{./figures/}}

\begin{document}

\title{Direct evaluation of the Force Constant Matrix in Quantum Monte Carlo}
\author{Y.Y.F.~Liu}
\email{ly297@cam.ac.uk}
\author{B.~Andrews}
\author{G.J.~Conduit}
\affiliation{Theory of Condensed Matter Group, Cavendish Laboratory, J. J. Thomson Avenue, Cambridge CB3 0HE, United Kingdom}
\date{\today}


\begin{abstract}
We develop a formalism to directly evaluate the matrix of force constants within a Quantum Monte Carlo calculation. We utilize the matrix of force constants to accurately relax the positions of atoms in molecules and determine their vibrational modes, using a combination of Variational and Diffusion Monte Carlo. The computed bond lengths differ by less than $0.007 \text{\AA}$ from the experimental results for all four tested molecules. For hydrogen and hydrogen chloride, we obtain fundamental vibrational frequencies within 0.1\% of experimental results and $\sim$10 times more accurate than leading computational methods. For carbon dioxide and methane, the vibrational frequency obtained is on average within 1.1\% of the experimental result, which is at least 3 times closer than results using Restricted Hartree-Fock and Density Functional Theory with a Perdew-Burke-Ernzerhof (PBE) functional and comparable or better than Density Functional Theory with a semi-empirical functional.
\end{abstract}

\maketitle


\section{Introduction}
\label{sec:Introduction}
Quantum Monte Carlo (QMC) is a leading class of approaches used to establish and study the electronic ground state of molecules and solids. Specifically, Diffusion Monte Carlo (DMC) is widely used to project out the exact electronic ground state wave function of a system, subject only to the fixed node approximation, fully accounting for correlation effects such as van der Waals interactions~\cite{wheatley2004intermolecular,wheatley2007calculating}. 
Although DMC is an ideal tool for studying the electronic wave function of the system~\cite{Kolorenc11}, the determination of the wave function of the atoms -- their expected positions and energy landscape -- remains a challenge for the method. Several approaches have been put forward to calculate the force acting on the atoms with DMC~\cite{Lee07, Badinski10, Badinski08, Moroni14, Filippi02ii, Assaraf00,Attaccalite08,filippi2016simple}, but a more comprehensive characterization of the atomic wave function requires the second derivative of the energy --  the matrix of force constants -- to both efficiently relax atomic positions and calculate vibrational modes.

We propose a method to directly calculate the matrix of force constants, $\diffd^2E/\diffd\vec{R}_{I}\diffd\vec{R}_{J}$, where $\vec{R}_{I}$ is the position of the $I$th, and $\vec{R}_{J}$ the $J$th, atom in the system. The energy, $E=\braket{\hat{H}}$, is calculated in the Born-Oppenheimer approximation of Hamiltonian $\hat{H}$; that is with the electrons always in their ground state for the respective atomic configuration. The calculation is implemented in QMC through a new quantum mechanical expectation value, $\diffd^2\langle\hat{H}\rangle/\diffd\vec{R}_{I}\diffd\vec{R}_{J}$, meaning that it can be evaluated with one configuration of the atoms to recover the entire matrix of force constants. The matrix of force constants allows us to efficiently relax atomic positions 
and determine the vibrational modes. 

We start by introducing the formalism and the QMC methods in \secref{sec:Formalism}. We subsequently outline the applications and implementation of the matrix of force constants in \secref{sec:AppFC}, followed by a series of case studies in~\secref{sec:CS}. We begin with atomic and diatomic hydrogen, and then move on to hydrogen chloride, carbon dioxide, and methane. For each molecule we derive the matrix of force constants, relax the positions of the atoms, and determine the vibrational modes. We critically evaluate the results with respect to existing computational methods: Restricted Hartree-Fock (RHF)~\cite{Hartree28, Slater28, Fock30} and Density Functional Theory (DFT)~\cite{Hohenberg64, Kohn65}. Finally, in \secref{sec:Discussion} we summarize the results and discuss future opportunities for the new formalism.


\section{Formalism}
\label{sec:Formalism}

In this section, we present the matrix of force constants. We then outline the numerics by discussing how the electronic orbitals are generated and the details of the QMC algorithms.

\subsection{Matrix of force constants}
\label{subsec:mfc}

We consider many-body quantum systems comprised of $N_\text{n}$ nuclei and $N_\text{e}$ electrons. The three-dimensional position vectors are denoted as $\mathbf{R}_I$ for nuclei and $\mathbf{r}_i$ for electrons, with $I=1,\dots,N_\text{n}$ and $i=1,\dots,N_\text{e}$. These are used to construct the corresponding multi-dimensional vectors in configuration phase space: $\mathbf{R}\equiv(\mathbf{R}_1,\dots,\mathbf{R}_{N_\text{n}})$ and $\mathbf{r}\equiv(\mathbf{r}_1,\dots \mathbf{r}_{N_\text{e}})$.

We use the non-relativistic Hamiltonian~\cite{Foulkes01}
\begin{eqnarray}
	\hat{H}=&-&\frac{1}{2}\sum_{i=1}^{N_\text{e}} \nabla_{\mathbf{r}_i}^2 + \sum_{i<j}^{N_\text{e}} \frac{1}{|\mathbf{r}_i - \mathbf{r}_j|} \nonumber \\
	&-& \sum_{i=1}^{N_\text{e}} \sum_{I=1}^{N_\text{n}} V_{I}(\mathbf{R}_I - \mathbf{r}_i) + \sum_{I < J}^{N_\text{n}} \frac{Z_I Z_J}{|\mathbf{R}_I - \mathbf{R}_J|}, \nonumber
\end{eqnarray}
which is comprised of the electron kinetic energy, as well as the electron-electron, electron-ion, and ion-ion interactions\footnote{We work with Hartree atomic units $\hbar=\text{m}_{\text{e}}=\text{e}=1$.}. $V_I$ and $Z_I$ represent the electron-ion pseudopotential and full nuclear charge, of atom $I$, respectively. 

We use a Hartree-Fock average effective Trail-Needs pseudopotential~\cite{Drummond16}, which has been specifically optimized for DMC calculations, to screen the effects of the core electrons and nucleus on the valence electrons.

In an electron position basis, the expectation value of the energy~\cite{Badinski10} may be written as
\begin{equation}
E=\frac{\int\Psi^*\hat{H}\Psi\dr}{\int|\Psi|^2\dr}\punc{,} \nonumber
\end{equation}
where the many-body wave function, $\Psi$, and Hamiltonian, $\hat{H}$, are both functions of nucleus configuration, $\vec{R}$, and electron configuration, $\vec{r}$.

The force acting on ion $\textit{I}$ is defined as the negative total derivative of the energy with respect to the nuclear coordinates. 
Taking the first derivative of the energy with respect to atom position~\cite{Badinski10} yields
\begin{equation}
	\frac{\diffd E}{\diffd\vec{R}_I}=\frac{\int\Psi^*\frac{\diffd\hat{H}}{\diffd\vec{R}_I}\Psi\dr}{\int|\Psi|^2\dr}+\Bigg[ \frac{\int\frac{\diffd\Psi^{*}}{\diffd\vec{R}_I}(\hat{H}-E)\Psi\dr}{\int|\Psi|^2\dr} +\text{c.c.} \Bigg]\punc{,} \nonumber
\end{equation}

which is decomposed into Hellmann-Feynman and Pulay terms, respectively.
When the wave functions are exact eigenstates of the Hamiltonian, such that $( \hat{H}-E )\Psi = 0 $, the Pulay term vanishes. However in practice, the wave functions are not exact in Variational Monte Carlo (VMC) or DMC, so the Pulay term needs to be included to obtain the total force. 

In this paper, we derive the matrix of force constants from the second derivative of the energy that takes the form of
\begin{align}
&\frac{\diffd^2E}{\diffd\vec{R}_I\diffd\vec{R}_J}
= \frac{\int\Psi^*\frac{\diffd^2\hat{H}}{\diffd\vec{R}_I\diffd\vec{R}_J}\Psi\dr}{\int|\Psi|^2\dr}  \nonumber \\
&\;\;\;\;\;\;\;\;\;+\frac{\int\Psi^*\Big[\frac{\diffd\Psi}{\diffd\vec{R}_I}\left(\Psi^{-1}\frac{\diffd\hat{H}}{\diffd\vec{R}_J}\Psi -\frac{\diffd E}{\diffd\vec{R}_J}\right)+(I\leftrightarrow J)\Big]\dr} {2\int|\Psi|^2\dr} \nonumber\\
&\;\;\;\;\;\;\;\;\;+\frac{\int \Big[\frac{\diffd}{\diffd\vec{R}_{J}} \left[ \frac{\diffd \Psi}{\diffd \vec{R}_I}\left( \hat{H}-E \right)\Psi \right] + (I\leftrightarrow J)\Big] \mathrm{d}\mathbf{r}}{2\int |\Psi|^2 \mathrm{d}\mathbf{r}}   +\text{c.c.} . \nonumber
\end{align}

This comprises one component of the matrix of force constants, so we must cycle over all atom pairs $\{I,J\}$ to determine the entire matrix.
The second derivative of the Hamiltonian with respect to atom position does not commute with the Hamiltonian, hence we approximate the pure expectation value of the force constants in the DMC procedure, as discussed in \secref{subsec:DMC}. The first two terms of the matrix of force constants stem from the Hellmann-Feynman force, whereas the third is due to the Pulay force.

To calculate the entire matrix of force constants using the Monte Carlo algorithm, we need to compute the ion-ion and electron-ion components at a cost of at most $O(N_\textrm{n}^3N_\textrm{e}) + O(N_\textrm{n}^2N_\textrm{e}^2)$. This leads to an overall dominant scaling of $O(N_\textrm{n}^4)$ assuming $O(N_\textrm{n}) \propto O(N_\textrm{e})$.

Having evaluated the matrix of force constants and implemented the formalism we can then use it to study atomic relaxation and vibrational modes.

\subsection{Variational Monte Carlo}
\label{subsec:vmc}

For the fermionic many-body trial wave function in the VMC method~\cite{McMillan65}, we take a Slater-Jastrow wave function of the form~\cite{Foulkes01}:
\begin{equation}
\Psi_\text{T}=\mathrm{e}^{J} D_{\uparrow}D_{\downarrow},\nonumber
\end{equation}
where $D_{\uparrow}(D_{\downarrow})$ denotes the Slater determinant of the molecular spin-up(down) orbitals. Here, the usual Hartree-Fock ansatz, $\Psi_{\text{HF}}=D_{\uparrow}D_{\downarrow}$, which encodes Pauli exclusion through the anti-symmetry of the Slater determinant, is multiplied by a Jastrow factor, $\text{e}^J$, which is an optimizable function used to impose further constraints on $\Psi_\text{T}$.

Initially, we compute the VMC energy, which is the expectation value of the Hamiltonian operator with respect to the trial wave function~\cite{Badinski10}:
\begin{equation}
E_\text{VMC}=\frac{\int |\Psi_\text{T}(\mathbf{r})|^2 E_\text{L}(\mathbf{r})\dr}{\int |\Psi_\text{T}(\mathbf{r})|^2\dr},\nonumber
\end{equation}
where $E_\text{L}=\Psi_\text{T}^{-1}(\mathbf{r})\hat{H}\Psi_\text{T}(\mathbf{r})$ is the local energy, $\dr$ is the infinitesimal hypervolume element in electron configuration phase space, and the integrals are performed using Monte Carlo\footnote{Specifically, the Metropolis algorithm is used to generate a set of configurations distributed according to the square modulus of a trial wave function over which the local energy is averaged.} in the CASINO program~\cite{Needs10}.

Single-particle orbitals for the different molecular structures were calculated using the CRYSTAL program~\cite{CRYSTAL14}. The RHF and DFT calculations with two exchange-correlation functionals the Perdew-Burke-Ernzerhof (PBE)~\cite{Perdew96} containing no exact orbital exchange and the B3LYP~\cite{becke1993density,stephens1994ab,vosko1980accurate} hybrid functional containing a fixed amount of exact exchange were performed with triple-$\zeta$-valence Gaussian basis sets, as well as polarization and diffuse basis functions~\cite{weigend2005balanced}.
The exact exchange-correlation functional is unknown and the choice of functional depends heavily on the system and the property of interest. PBE as a general functional was chosen for its greater predictive power across all simulations and properties~\cite{ernzerhof1999assessment}, though it is less likely to achieve the accuracy of a semi-empirical functional such as B3LYP, which was chosen in addition for its good agreement with post-DFT methods within its range of applicability on molecules~\cite{mardirossian2017thirty,paier2007does}.
We use a Jastrow factor in its most general form comprising of an electron-electron term, an electron-nucleus term, and an electron-electron-nucleus term.
The wave function parameters are optimized by using the variance minimization method~\cite{Drummond05ii} first, followed by the energy minimization method~\cite{Umrigar88,Umrigar07,brown2007energies}. 

\subsection{Diffusion Monte Carlo}
\label{subsec:DMC}

DMC evolves a wave function, $\Phi$, according to the imaginary-time Schr\"{o}dinger equation, in order to project out the lowest energy eigenstate, $\Phi_0$, with the same nodal surface as the trial wave function~\cite{Foulkes01}.

The efficiency of the DMC algorithm is improved by importance sampling~\cite{Ceperley80}. By multiplying the wave function, $\Phi$, by a trial wave function, $\Psi_\text{T}$ from VMC, we may solve the Schr\"{o}dinger equation for the mixed distribution $f(\mathbf{r},\tau)=\Phi(\mathbf{r},\tau)\Psi_\text{T}(\mathbf{r})$, where $\tau$ denotes imaginary time. We tested, and found negligible error, with time-steps of $\tau=0.01~\text{a.u.}$, and so this is used throughout~\cite{Lee11}. The fixed-node approximation~\cite{Anderson75,Anderson76} is introduced to overcome the fermion sign problem by constraining the nodal surface of $\Phi_0$ to match that of $\Psi_\text{T}$~\footnote{The fixed-node approximation is the only uncontrolled approximation in a DMC simulation of all-electron systems.}.

The expectation value of the energy in the DMC method~\cite{Badinski10} is given by
\begin{equation}
E_\text{DMC}=\frac{\int\Phi(\mathbf{r})\Psi_\text{T}(\mathbf{r})E_\text{L}(\mathbf{r})\dr}{\int \Phi(\mathbf{r})\Psi_\text{T}(\mathbf{r}) \dr}. \nonumber
\end{equation}
This is an unbiased estimator, up to the approximations made, since $E_\text{DMC}$ does not depend on the trial wave function used. However, the mixed expectation value of an operator that does not commute with the Hamiltonian is biased. In these cases, we approximate the pure expectation value of an operator $\hat{O}$ with the extrapolation formula~\cite{ceperley1979monte} 
\begin{equation}
O=2O_\text{DMC}-O_\text{VMC}+\mathcal{O}\left[ (\Phi-\Psi_\text{T})^2 \right]. \nonumber
\end{equation}
Alternatively, the future-walking method may be used, for example, to obtain an exact pure estimator~\cite{Barnett91}.
Although the extrapolation formula improves the results, this procedure depends on an almost complete error cancellation and is strongly dependent on the quality of the wave function employed. We run the simulations for longer to systematically reduce the statistical error associated with variational techniques.

\begin{figure}
	\begin{minipage}[b]{\linewidth}
		\begin{minipage}[b]{.517\linewidth}
			\centering
			\begin{tikzpicture}
			\node at (0,0) {\centering\includegraphics{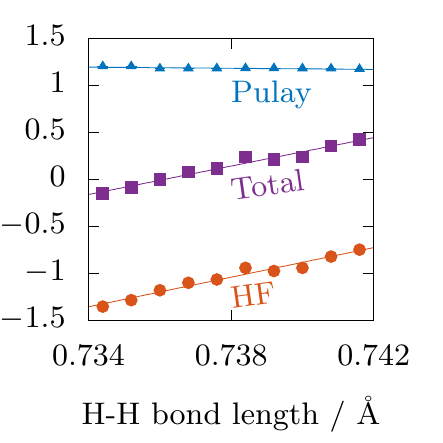}};
			\node[overlay] at (-2.5,2.05) {(a)};
			\label{fig:H2-pulay}
			\end{tikzpicture}
		\end{minipage}%
		\begin{minipage}[b]{.517\linewidth}
			\begin{tikzpicture}
			\node at (0,0) {\centering\includegraphics{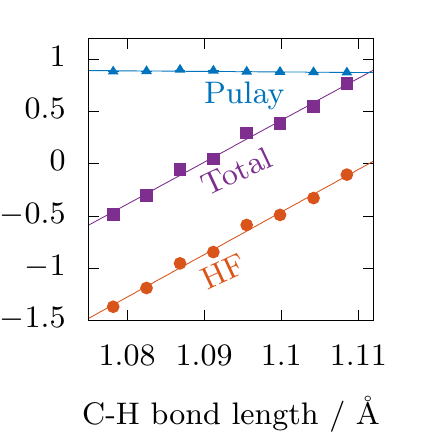}};
			\node[overlay] at (-2.4,2.05) {(b)};
			\label{fig:CH4-pulay}
			\end{tikzpicture}
		\end{minipage}%
	\end{minipage}
	\caption{Interatomic force estimates for the molecules (a) $\text{H}_{2}$ and (b) $\text{CH}_{4}$. The red dots correspond to the Hellmann-Feynman (HF) force, the blue triangles to the Pulay force, and the magenta squares to the total force. Error bars for all of the data are given -- some error bars are smaller than the data points on the scale of the plot.}
	\label{fig:H2-CH4-pulay}
\end{figure}

\subsection{Contributions of the Hellmann-Feynman and Pulay terms}
\label{subsec:hf_pulay}

Both the Hellmann-Feynman and Pulay terms contribute to the force and matrix of force constants, and both Pulay terms are zero at the exact electronic ground state. However, the Pulay contribution to the matrix of force constants contains a second derivative of the wave function with respect to atom position, and so is more susceptible to steep gradients due to an incorrect trial wave function. Therefore, when using the electronic structure methods, it is useful to determine the relative contributions of the Hellmann-Feynman and Pulay terms so that we can gauge the importance of refining the electronic wave function. 

The interatomic force in a hydrogen molecule and methane molecule is shown in Fig.~\ref{fig:H2-CH4-pulay}. Different bond lengths within the vicinity of the equilibrium were chosen and forces were evaluated using the methods described in Secs.~\ref{subsec:vmc} and~\ref{subsec:DMC}. The addition of the Pulay force to the Hellmann-Feynman force shifts the equilibrium bond length by 2\% in both examples, therefore the Pulay force is crucial for finding the correct equilibrium geometry. 

We now turn to consider the calculation of the matrix of force constants -- the gradient of the force. We first note from Fig.~\ref{fig:H2-CH4-pulay} that the Pulay force is remarkably constant with respect to bond length across all molecules tested in this paper, regardless of the molecular geometry. This means that the gradient of the force is negligible, and therefore does not significantly contribute to the matrix of force constants. We find that, when directly evaluated, the value of the Pulay term in the matrix of force constants is smaller than its standard error, as well as the standard error of the contribution from the Hellmann-Feynman term. Furthermore, based on the analysis of variance at the $\alpha = 0.05$ level, the gradient of the Pulay force does not significantly deviate from zero. This means that the change in vibrational frequency due to the Pulay force gradient is just 1\% of that from the Hellmann-Feynman gradient for both hydrogen and methane. This conclusion is also backed up by independent studies: taking a numerical derivative of the results for $\text{H}_2$ and $\text{LiH}$ reported by Casalegno et al.~\cite{Casalegno03}, $\text{CO}_2$ reported by Lee et al.~\cite{Lee07}, as well as adenine-thymine reported by Ruiz-Serrano et al.~\cite{ruiz2012pulay}, confirms the small contribution of the Pulay term to the matrix of force constants. 

The effect of the Pulay term in the force is significant for force analysis and, when suitably formulated, can reduce the statistical noise in the expectation value and improve the convergence of the optimization algorithm~\cite{ruiz2012pulay}. However, as the Pulay force is almost constant with interatomic bond length, its contribution to the matrix of force constants is negligible. Therefore, we expect the Pulay contribution to the matrix of force constants to be insensitive to the quality of the trial wave function. 
Another corollary is that for our zero-variance scheme~\cite{Assaraf99,Assaraf03}, the expected $-5/2$ power law tail associated with infinite variance that could arise from Pulay terms~\cite{Trail08i,Badinski10} will only make a limited contribution to the tail of the total probability distribution and as shown in Fig.~\ref{fig:H2_histogram} we have not been hindered by this problem in our practical application. 


\section{Applications of Force Constants}
\label{sec:AppFC}

\subsection{Atomic relaxation}
\label{subsec:atomic_relax}

The primary requirement for a versatile geometry model, is the ability to minimize the energy of an arbitrary configuration of atoms~\cite{Wagner10}. For quantum mechanical simulations, this is often performed using VMC, due to the algorithm's efficiency. 
In this paper, we relax the positions of the atoms first with VMC using the additional information provided by the matrix of force constants. The wave function from VMC is then optimized in DMC and we perform the same iteration steps using DMC to confirm convergence and further reduce the error~\cite{Needs10,al2008variational,barnett1993variational,drummond2006quantum}.

Requiring that the energy of the system is constant up to quadratic order in atomic displacement, and explicitly correcting for global translation and rotation, as well as anharmonicity, we find that the atomic displacement, $\Delta\mathbf{R}$, is given by
\begin{equation}
\Delta\mathbf{R} = -2 \mathbf{M}^{-1} \cdot \nabla_\mathbf{R} E - \frac{\sum_I m_I \mathbf{R}_I}{\sum_I m_I}-\mathbf{R}\times\boldsymbol{\theta},
\label{eq:displacement}
\end{equation}
where $\mathbf{M}\equiv\diffd^2E/\diffd\vec{R}_{I}\diffd\vec{R}_{J}$ is the matrix of force constants; $\nabla_\mathbf{R} E\equiv \diffd E/\diffd\vec{R}_{I}$ is the multi-particle gradient of the energy with respect to atom position; $\boldsymbol{\theta}$ is the three-dimensional global angular displacement vector for the configuration $\mathbf{R}$. On each step, we displace the atoms by $\Delta\mathbf{R}$ in order to compare with other methods in determining the minimum energy of the system. This yields the interatomic bond length at the minimum of the total potential. The details of the calculation are outlined in Appendix~\ref{sec:details_ar}.

Though $\langle\hat{\vec{R}}\rangle$ minimizes the total potential after the atomic relaxation procedure, if the potential is not symmetric then the expected separation of the atoms does not coincide with the minimum. We capture the lowest-order difference with the addition of an anharmonic correction term $\Delta\mathbf{R}_\text{a}$, outlined in Appendix~\ref{subsec:corr_anharm}.

\subsection{Vibrational modes}
\label{subsec:vibrational_modes}

One main motivation for incorporating the matrix of force constants into the QMC procedure is the ability to calculate vibrational modes and frequencies directly. In this paper, we use a variety of methods to calculate the vibrational frequency for a cross comparison.

Up to a simple mapping, the eigenvectors of the matrix of force constants correspond to the vibrational modes of the system, and the eigenvalues correspond to the vibrational frequencies. We can, therefore, use a complete diagonalization of the matrix of force constants to estimate the eigenmodes and frequencies. To benchmark the results, we also calculate the frequencies from a numerical second derivative of the energy with respect to bond length -- referred to as the energy curvature (EC) method -- and from a numerical derivative of the force -- force gradient (FG) method. 

A discussion of all of these methods, including the analysis of statistical uncertainty and the anharmonic correction, is detailed in Appendix~\ref{sec:details_vm}. 


\section{Case Studies}
\label{sec:CS}

In this section, we evaluate the effectiveness of the matrix of force constants formalism for a selection of molecules. We first confirm the theory with the simplest possible molecules, before testing the generalizability of the formalism on molecules containing more atoms.

\subsection{Hydrogen atom and molecule}
\label{subsec:H}

We begin by analyzing the simplest physical system: the hydrogen atom. By performing a DMC calculation, we verify that the hydrogen atom obeys Newton's laws since it has a net force of $(3.68\pm5.17)\times10^{-3}~E_\text{h}\text{\AA}^{-1}$ acting on it, which is zero within standard error. Furthermore, the hydrogen atom has a computed eigenfrequency of $0~\text{cm}^{-1}$. This system behaves as expected and confirms the translational invariance.

From this, it is natural to increment the complexity by adding another hydrogen atom to form an $\mathrm{H}_2$ molecule. This is the simplest physical example that allows us to verify the eigenfrequencies from our DMC method, which has no nodes and gives an exact wave function, by comparing them against both experimental results in the literature, and RHF/DFT predictions from the CRYSTAL program. 
\begin{figure}
\begin{tikzpicture}
\node at (0,0) {\centering\includegraphics{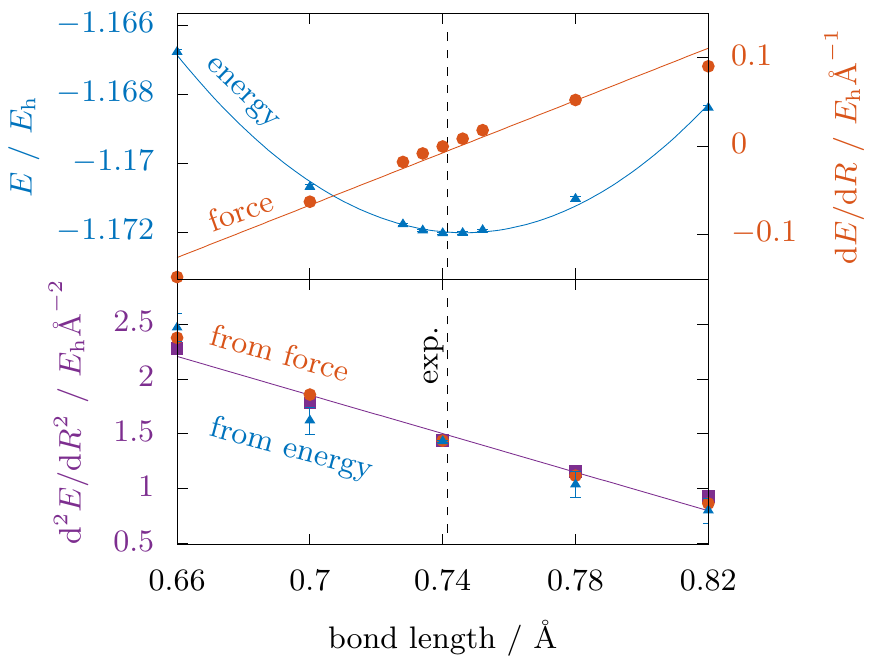}};
\node[overlay] at (-4.5,3.2) {(a)};
\node[overlay] at (-4.5,0.5) {(b)};
\end{tikzpicture}
\caption{(a) Energy, force, and (b) diagonal force constant against bond length, for the hydrogen molecule. For the energy and force plots, the parabolic and linear curves of best fit for the visible data, are overlaid. For the force constant plots, we show the diagonal force constant derived using the finite difference method from the energy curvature, the force gradient, and the direct analytical evaluation of the force constant. The line of best fit for the visible MFC data is overlaid. The dashed line indicates the experimental equilibrium bond length. Error bars for all of the data are given -- some error bars are smaller than the data points on the scale of the plot.}
\label{fig:H2_second_derivative}
\end{figure}

\begin{figure}
\includegraphics{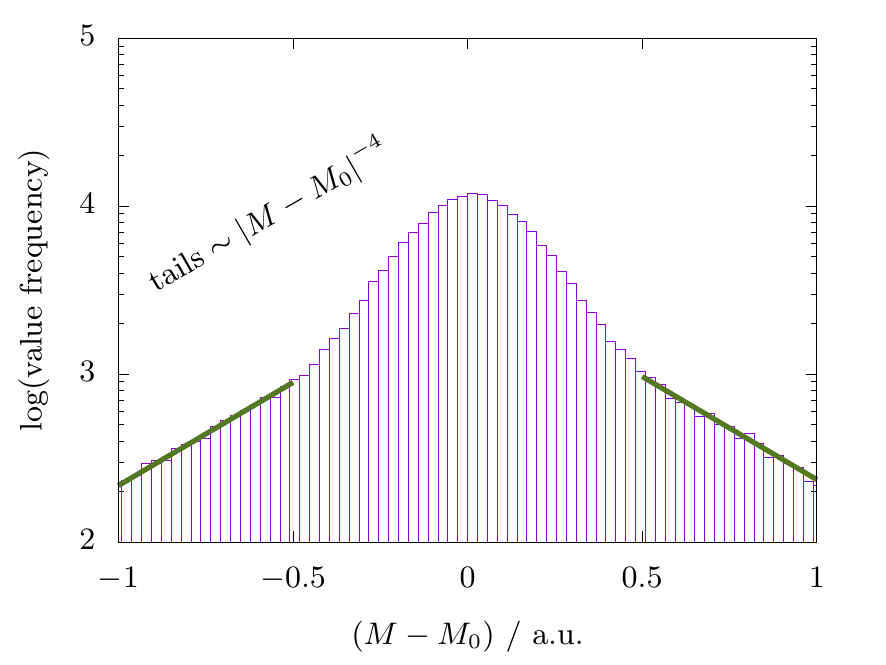}
\caption{Distribution of probability densities for the observed values of an element of the matrix of force constants from DMC, offset by the mean $M_0$. Data are shown for the hydrogen molecule at the equilibrium bond length.}
\label{fig:H2_histogram}
\end{figure}

The energy, force, and diagonal elements of the matrix of force constants for the hydrogen atom is shown as a function of bond length in Fig.~\ref{fig:H2_second_derivative}. We verify that the energy is at a minimum and the force is zero at the correct equilibrium bond length of $74.13~\text{pm}$~\cite{Sutton65}, within standard error.  Furthermore, in Fig.~\ref{fig:H2_second_derivative}b, we show that in the vicinity of the equilibrium bond length for the hydrogen molecule, the energy curvature, the force gradient, and the direct computation, all agree within error bounds. The entire matrix of force constants is sparse. If we are only interested in the vibrational mode, it can be reduced to be a $2\times2$ matrix, with the off-diagonal force constants to be minus the diagonal elements within error, as required by symmetry~\footnote{The diagonal elements of the matrix of force constants are the same for the hydrogen molecule, and hence no particular atom is specified when referring to $\mathrm{d}^2E/\mathrm{d}R^2$ in Figs.~\ref{fig:H2_second_derivative} and \ref{fig:H2_histogram}.}. Note that here the diagonal elements of the matrix of force constants are not constant across the range of bond lengths shown, as can also be seen in the slight curvature of the force in Fig.~\ref{fig:H2_second_derivative}a. This is due to the anharmonicity of the potential in a diatomic molecule. We may use the gradient of the force constant to calculate the anharmonic constant, and correct for the anharmonicity, as discussed in Appendix~\ref{subsec:corr_anharm}.

For the hydrogen molecule, it is possible to extract the matrix of force constants efficiently from the force, or energy, because the computational cost of obtaining the numerical derivatives is low. 
However, for more complicated molecules, where structural optimization is influential, using the matrix of force constants would be beneficial, as it provides both the movement direction and amplitude towards the minimum energy configuration. In these cases, the equivalent information would take considerably longer to extrapolate from either energy or force, if possible.

\begin{figure}
\includegraphics{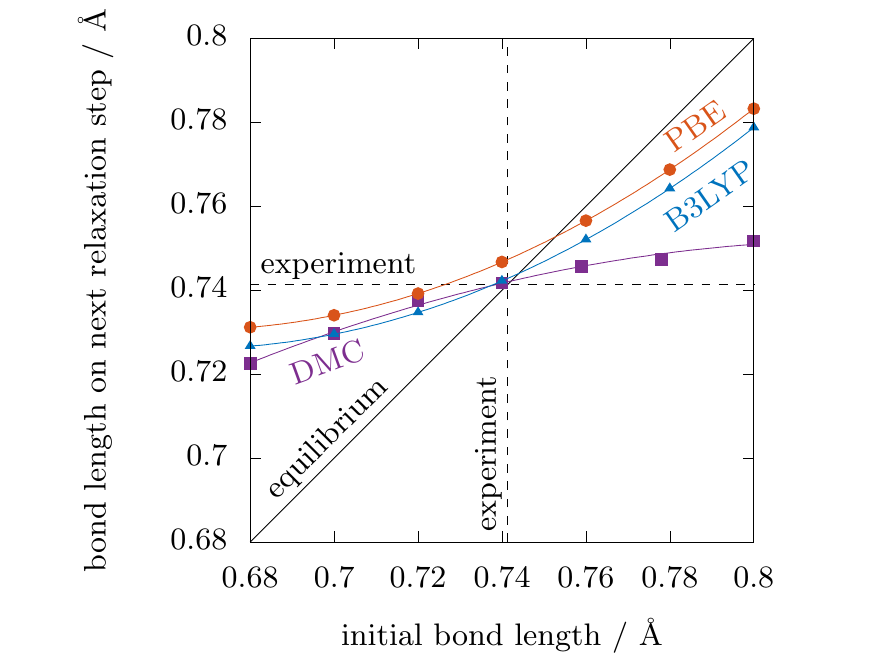}
\caption{Bond length on next step of atomic relaxation as a function of initial bond length, for the hydrogen molecule. The unconstrained parabolas of best fit, with respect to the visible DFT (PBE, B3LYP) and DMC data points, are overlaid. The dashed lines indicate the experimental equilibrium bond length, whereas the solid line indicates the equilibrium fixed points with respect to the plot. DMC error bars are smaller than the data points on the scale of the plot.}
\label{fig:H2_convergence}
\end{figure}

Equipped with reliable results for the matrix of force constants directly from DMC at each bond length, we may now exploit this information to efficiently relax the bond length of the molecule. The force tells us the direction to move the atoms, and the matrix of force constants additionally tells us how far to move them, on each step (Eq.~\ref{eq:displacement}). Owing to the anharmonicity of the potential, we must relax to the equilibrium bond length of $74.13~\text{pm}$ over several steps. The predicted bond length on the next atomic relaxation step as a function of initial bond length is shown in Fig.~\ref{fig:H2_convergence}. We see that the PBE curve intersects the equilibrium line at $0.753~\text{\AA}$ and the B3LYP curve intersects at $0.745~\text{\AA}$ ; whereas our DMC calculation intersects at $0.7420\pm0.0007~\text{\AA}$, in close agreement with the experimental value of $0.74130~\text{\AA}$~\cite{Sutton65}\footnote{Note that the turning points of the PBE, B3LYP and DMC curves in Fig.~\ref{fig:H2_convergence} do not correspond to fixed points, but rather the fixed points are given by the intersections of the curves with the equilibrium line.}. Furthermore, we note that PBE and B3LYP curves have a similar shape as a result of sharing the same optimization algorithm. However, both are steeper than the DMC curve in the vicinity of equilibrium and therefore converge more slowly, due to the fact that the DFT algorithm uses an inaccurate estimate for the force constant. The number of iteration steps reduced are particularly apparent when there are multiple atoms in a molecule, however the lower number of steps does not necessarily indicate a more efficient algorithm, as the complexity of each step needs to be taken into consideration. In some cases where complex molecules cannot be relaxed sufficiently for a long time using DFT, our approach may be more efficient in giving the ground state geometry. In general, due to DFT's inaccurate estimate of the force constant, we observe at least a slight improvement for all molecules.
\begin{table*}
	\begin{ruledtabular}
		\begin{tabular}{c c c c c c c c}
			\text{mode} & $\omega_\text{RHF}$ & $\omega_\text{PBE}$ & $\omega_\text{B3LYP}$ & $\omega_\text{EC}$ & $\omega_\text{FG}$ & $\omega_\text{DMC}$ & $\omega_{\text{exp}}$\cite{Hydrogen} \\
			\hline
			\text{stretch} & $4379$ & $4116$ & $4384$ & $4170\pm10$ & $4180\pm8$ & $4166\pm 4$ & $4161.1663\pm0.0002$
		\end{tabular}
	\end{ruledtabular}
	
	\caption{\label{tab:H2}Vibrational frequencies, evaluated at the computed equilibrium bond length, for the hydrogen molecule in units of $\text{cm}^{-1}$, where $\omega_{\text{RHF}}$ denotes the vibrational frequency obtained from a RHF calculation, $\omega_\text{PBE}$ and $\omega_\text{B3LYP}$ from DFT calculations with a PBE /B3LYP functional, $\omega_\text{EC}$ from the curvature of the DMC energy, $\omega_\text{FG}$ from the gradient of the DMC force, $\omega_\text{DMC}$ from the DMC matrix of force constants, and $\omega_\text{exp}$ from experiment. All values are presented at zero temperature and post anharmonic corrections. These quantities, as well as their associated errors, are discussed in Appendix~\ref{sec:details_vm}.}
\end{table*}

An additional important check, before we proceed, is an analysis of the probability distribution of the matrix of force constants generated by DMC. Fig.~\ref{fig:H2_histogram} shows the histogram of a force constant value for the DMC run at the computed equilibrium bond length. From this we can see that the force constant heavy tails decay with the same $\sim|M-M_0|^{-4}$ power law as the energy and Hellmann-Feynman force~\cite{Trail08i, Badinski10}. This is as expected, since the effective remaining term is the Hellmann-Feynman term due to the quasi-constant Pulay contribution in proximity to the ground state, as seen in \secref{subsec:hf_pulay}.
Reassuringly, the expected value of the distribution is also the modal value.

Now that the configuration is relaxed, we may analyze the fundamental vibrational modes. For the hydrogen molecule, we obtain six eigenmodes, as expected for a diatomic molecule. Three modes correspond to global translation, two correspond to global rotation, and one corresponds to a symmetric stretch. The symmetric stretch mode has the largest eigenfrequency. We extract the frequency using a selection of methods, outlined in Appendix~\ref{sec:details_vm}, for a cross-comparison. In this case, we obtain a fundamental vibrational frequency of $\omega_\text{DMC}=4166\pm 4~\text{cm}^{-1}$, compared to the experimental value of $\omega_\text{exp}=4161.1663\pm0.0002~\text{cm}^{-1}$, which is $4.83~\text{cm}^{-1}$ away. This is a firm statistical confirmation of the accuracy of DMC compared to RHF, PBE and B3LYP results, which have deviations of  $218~\text{cm}^{-1}$,  $45~\text{cm}^{-1}$ and $223~\text{cm}^{-1}$ from experiment, respectively. All of our computational values for the vibrational frequency from DMC -- matrix of force constants, force gradient, and energy curvature -- agree with each other within standard error and show close agreement to experiment. The DMC procedure yields no translational or rotational modes, just as for the hydrogen atom. A summary of the results is shown in Table~\ref{tab:H2}. Note that the less computationally expensive calculation of the energy was run for four times longer, compared to the force gradient and matrix of force constant methods, in order to give the error bars of the energy curvature eigenfrequency to a comparable value.

\subsection{Hydrogen chloride}
\label{subsec:HCl}

Now that we have verified that the matrix of force constants agrees with numerical estimates, and that by exploiting this information it is possible to relax the molecule more efficiently, and outperform RHF and DFT estimates for the fundamental vibrational frequency for the hydrogen case, we move onto a more complex molecule: hydrogen chloride. We increment the complexity of our case study in order to verify that our formalism can cope with an asymmetric system with unequal masses.

The hydrogen chloride molecule again relaxes quickly to equilibrium, with a computed bond length of $128.0\pm0.6~\text{pm}$, which agrees with the experimental value of $127.5~\text{pm}$ within standard error. Both atoms have the appropriate displacement to ensure that the center of mass is stationary. We obtain six eigenmodes for the system, including one symmetric stretch mode with eigenfrequency $\omega_\text{DMC}=2995\pm 8~\text{cm}^{-1}$. This result agrees with the experimental value of $\omega_\text{exp}=2990.946\pm0.003~\text{cm}^{-1}$ within standard error, whereas RHF, PBE, B3LYP methods are $107~\text{cm}^{-1}$,  $112~\text{cm}^{-1}$, $50~\text{cm}^{-1}$ away, respectively. A summary of the results is shown in Table~\ref{tab:HClandCO2andCH4}a. 
\begin{table}
(a) hydrogen chloride
\vspace{0.5em}
\begin{ruledtabular}
\begin{tabular}{c c c c c c}
\text{mode} & $\omega_\text{RHF}$ & $\omega_\text{PBE}$ & $\omega_\text{B3LYP}$ & $\omega_\text{DMC}$ & $\omega_{\text{exp}}$\cite{Diatomics} \\
\hline
\text{stretch} & $3098$ & $2879$ & $2941$ & $2995\pm 8$ & $2990.946\pm0.003$
\end{tabular}
\end{ruledtabular}
\vspace{0.5em}
(b) carbon dioxide
\vspace{0.5em}
\begin{ruledtabular}
\begin{tabular}{c c c c c c}
mode & $\omega_\text{RHF}$ & $\omega_\text{PBE}$ & $\omega_\text{B3LYP}$ & $\omega_\text{DMC}$ & $\omega_{\text{exp}}$\cite{Molecules} \\
\hline
sym. stretch & $1468$ & $1284$ & $1325$ & $1309\pm 5$ & $1333\pm6$ \\
antisym. stretch & $2480$ & $2297$ & $2321$ & $2312\pm 6$ & $2349\pm1$ \\
bending & $766$ & $634$ & $664$ & $662\pm 2$ & $667\pm1$ \\
\end{tabular}
\end{ruledtabular}
\vspace{0.5em}
(c) methane
\vspace{0.5em}
\begin{ruledtabular}
\begin{tabular}{c c c c c c}
mode & $\omega_\text{RHF}$ & $\omega_\text{PBE}$ & $\omega_\text{B3LYP}$ & $\omega_\text{DMC}$ & $\omega_{\text{exp}}$\cite{Molecules} \\
\hline
sym. stretch & $3101$ & $3034$ & $3074$ & $2874\pm8$ & $2917\pm1$ \\
scissor & $1655$ & $1496$ & $1544$ & $1534\pm 9$ & $1534\pm1$
\end{tabular}
\end{ruledtabular}
\vspace{1em}
\caption{\label{tab:HClandCO2andCH4} Vibrational frequencies, evaluated at the computed equilibrium bond length, for (a) the hydrogen chloride, (b) carbon dioxide, and (c) methane, molecules in units of $\text{cm}^{-1}$, where $\omega_{\text{RHF}}$ denotes the vibrational frequency obtained from a RHF calculation, $\omega_\text{PBE}$ and $\omega_\text{B3LYP}$ from DFT calculations with a PBE functional and a B3LYP hybrid functional respectively, $\omega_\text{DMC}$ from a DMC calculation, and $\omega_\text{exp}$ from experiment. All values are presented at zero temperature and post anharmonic corrections. These quantities, as well as their associated errors, are discussed in Appendix~\ref{sec:details_vm}.}
\end{table}

\subsection{Carbon dioxide}
\label{subsec:CO2}

\begin{table}
	\begin{ruledtabular}
		\begin{tabular}{c c c c c c}
			molecule & $x_0^\text{RHF}$ & $x_0^\text{PBE}$ & $x_0^\text{B3LYP}$ & $x_0^\text{DMC}$ & $x_0^{\text{exp}}$~\cite{Sutton65} \\
			\hline
			$\text{H}_2$ & $0.736$ & $0.753$ & $0.745$ & $0.7420\pm0.0007$ & $0.74130$ \\
			$\text{HCl}$ & $1.260$ & $1.286$ & $1.278$ & $1.280\pm0.006$ & $1.275$ \\
			$\text{CO}_2$ & $1.145$ & $1.182$ & $1.171$ & $1.167\pm0.003$ & $1.1598$ \\
			$\text{CH}_4$ & $1.089$ & $1.104$ & $1.098$  & $1.097\pm0.002$ & $1.093$ \\
		\end{tabular}
	\end{ruledtabular}
	\caption{Computed equilibrium bond lengths for the hydrogen, hydrogen chloride, carbon dioxide, and methane molecules, in units of \AA, where $x_0^{\text{RHF}}$ denotes the equilibrium bond length obtained from a RHF calculation, $x_0^\text{PBE}$ and $x_0^\text{B3LYP}$ from DFT calculations with a PBE/B3LYP functional, $x_0^\text{DMC}$ from a DMC calculation, and $x_0^\text{exp}$ from experiment. The details of the atomic relaxation calculation in DMC are discussed in Appendix~\ref{sec:details_ar}.}
	\label{tab:bonds}
\end{table}

In the previous two examples, we found that the matrix of force constants can correctly calculate the modes of a diatomic molecule. Building on this, we increment the complexity to carbon dioxide: a three-atom system with several non-trivial vibrational modes, some of which are in orthogonal directions.

In this case, the $\mathrm{O=C=O}$ configuration is relaxed to an equilibrium $\mathrm{C=O}$ bond length of $116.7\pm0.3~\text{pm}$ along one axis, which is within three standard deviations of the experimental value of $115.98~\text{pm}$. For carbon dioxide, we obtain nine vibrational modes: three of which correspond to global translations, two to global rotations, and four to vibrational modes. Of the vibrational modes, we obtain one symmetric stretch mode, one asymmetric stretch mode, and two bending modes along orthogonal axes.

The modes examined in this section show a consistent improvement over the RHF and PBE calculations, with DMC eigenfrequency deviations from experiment of $-1.80$\% (symmetric), $-1.58$\% (antisymmetric), and $-0.75$\% (bending). The recovery of the non-trivial antisymmetric mode is our first example to break the underlying symmetry of the molecule, and the bending mode shows that our formalism can extend to atoms moving in orthogonal directions. 
On average, our DMC result is $22~\text{cm}^{-1}$ away from the experimental value, which is an improvement over the RHF results (on average $122~\text{cm}^{-1}$ away) and PBE results (on average $45~\text{cm}^{-1}$ away). 
We note that in this particular case the B3LYP results are on average $13~\text{cm}^{-1}$ away from the experimental results, which is why it is a popular choice for non metal-containing molecules~\cite{peverati2011communication}.

It is worthwhile to mention that the experimental results come with a larger error for carbon dioxide when compared to smaller molecules, as shown in Table~\ref{tab:HClandCO2andCH4}b. The symmetric stretch mode is Raman active and infrared inactive, whereas for the other modes, the opposite is true~\cite{Molecules}. The Raman measurement typically comes with a larger uncertainty than infrared spectroscopy in this case, complicating the comparison to DMC. Additionally, for these larger molecules, as the number of modes increases, the chances of eigenfrequency interference is increased. Here we notice that the symmetric stretch mode eigenfrequency is quasi-degenerate with twice the bending mode eigenfrequency in Table~\ref{tab:HClandCO2andCH4}b, which could also potentially contribute to the increased uncertainty of the symmetric stretch mode~\cite{McCluskey06}. Finally, we note that the precise eigenfrequencies for arbitrarily large molecules have not been studied as extensively. In contrast, the eigenfrequency calculation for hydrogen especially, as well as for other common diatomic molecules, is often used as an experimental benchmark~\cite{Hydrogen}. Together these factors motivate the need for improving the accuracy and precision of electronic structure calculations, such as QMC.

\subsection{Methane}
\label{subsec:CH4}

For the last example, we extend our formalism to a three-dimensional molecule containing five atoms: methane, to demonstrate that the formalism can be applied to diverse configurations of atoms.
\begin{figure}
\begin{minipage}[b]{\linewidth}
\begin{minipage}[b]{.5\linewidth}
\begin{tikzpicture}
\node at (0,0) {\centering\includegraphics[width=\linewidth]{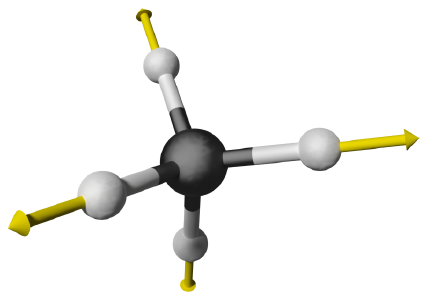}};
\node[overlay] at (-2,1.7) {(a)};
\label{fig:A1}
\end{tikzpicture}
\end{minipage}%
\begin{minipage}[b]{.5\linewidth}
\begin{tikzpicture}
\node at (0,0) {\centering\includegraphics[width=\linewidth]{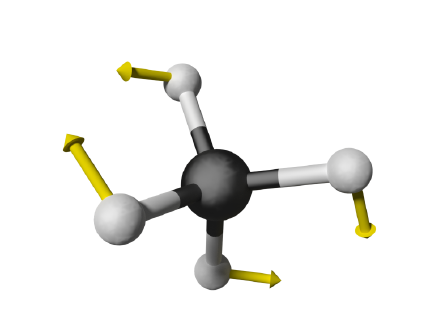}};
\node[overlay] at (-2,1.7) {(b)};
\label{fig:E}
\end{tikzpicture}
\end{minipage}%
\end{minipage}
\caption{(a) The symmetric stretch, and (b) scissor, vibrational modes of methane.}
\label{fig:CH4}
\end{figure}

We find that the configuration relaxes to a $\mathrm{C-H}$ bond length of $109.7\pm0.2~\text{pm}$, within two standard deviations of the experimental value of $109.3~\text{pm}$, in fewer iterations than existing methods. The equilibrium bond lengths for all case studies are summarized in Table~\ref{tab:bonds}. In this case, we obtain fifteen eigenmodes of the system: three corresponding to global translation, three corresponding to global rotation, and nine corresponding to non-trivial vibrational modes. Of the vibrational modes, we select two modes to examine in detail: the symmetric stretch mode and a scissor mode, as summarized in Table~\ref{tab:HClandCO2andCH4}c and illustrated in Fig.~\ref{fig:CH4}.

An analysis of these modes yields a DMC deviation from experiment of $-1.47$\% for the symmetric stretch mode and an expected agreement for the scissor mode, which is generally comparable to the results for carbon dioxide i.e. still of the order of $1$\% from the experimentally measured values. The successful recovery of these modes demonstrates that the formalism holds in three dimensions, and the excellent agreement for the scissor mode demonstrates that we are able to capture a non-trivial symmetry for this molecule. The symmetric stretch DMC eigenfrequency is $43~\text{cm}^{-1}$ away from experiment, whereas the RHF, PBE and B3LYP results are  $184~\text{cm}^{-1}$,  $117~\text{cm}^{-1}$ and  $157~\text{cm}^{-1}$ away, respectively.


\section{Conclusion}
\label{sec:Discussion}

In this paper, we develop and implement a formalism to evaluate the matrix of force constants in QMC. We calculate vibrational frequencies and improved estimates for the atomic displacements on each relaxation step, as well as correcting for anharmonicity. We report statistically significant improvements over RHF and DFT methods in the vast majority of cases, both in terms of the vibrational frequency and the efficiency of the atomic relaxation, for the hydrogen, hydrogen chloride, carbon dioxide, and methane molecules.     

The ability to calculate the matrix of force constants within DMC, in particular, makes us well-positioned to calculate vibrational modes where high accuracy is a necessity and relax atomic positions in complex systems with many degrees of freedom where the extrapolation from energy or force is difficult, if not impossible, to optimize the geometry. The approach applies to both molecules and periodic configurations. This will be especially beneficial in systems with heavy atoms that are challenging to analyze accurately with DFT, systems with significant anharmonic corrections, and also those with strong van der Waals interactions, such as layered materials and surfaces.


\begin{acknowledgments}
The authors thank Andrew Fowler, Neil Drummond, Christian Carbogno, Ryan Hunt, Pablo~L\'opez~R\'ios, John Trail, Thomas Gasenzer, and Lars Schonenberg for useful discussions. The QMC calculations in this paper were performed using {\sc CASINO}. Y.Y.F.L. acknowledges support from the Agency for Science, Technology and Research; B.A. acknowledges support from the Engineering and Physical Science Research Council under grant EP/M506485/1; and G.J.C. acknowledges support from the Royal Society. There is open access to this paper and data available at \texttt{https://www.openaccess.cam.ac.uk}.
\end{acknowledgments}


\appendix

\section{Atomic relaxation calculation}
\label{sec:details_ar}

In this section, we describe in detail how the configuration coordinates are adjusted on each step during the atomic relaxation process.

Let us define the atomic displacement on each Monte Carlo step as 
\begin{equation*}
\Delta\mathbf{R}=\Delta\mathbf{R}_{\text{e}}+\Delta\mathbf{R}_{\text{t}}+\Delta\mathbf{R}_{\text{r}},
\end{equation*}
where $\Delta\mathbf{R}_{\text{e}}$ is the energy-minimizing term, $\Delta\mathbf{R}_{\text{t}}$ is the correction for global translations, $\Delta\mathbf{R}_{\text{r}}$ is the correction for global rotations. We adjust the atomic displacements from $\mathbf{R}$ to $\mathbf{R}+\Delta\mathbf{R}$ on each step, so as to minimize the total energy of the system. Once the equilibrium is reached, the anharmonic correction $\Delta\mathbf{R}_{\text{a}}$ is applied.

\subsection{Minimizing the energy}

Consider a system of $N_\text{n}$ atoms in three dimensions. Taylor expanding the total energy of the system as a function of atomic displacements, up to quadratic order, yields
\begin{equation*}
E=E_0 + \sum_{I=1}^{N_\text{n}} \frac{\mathrm{d}E}{\mathrm{d}\mathbf{R}_I}\Delta \mathbf{R}_I + \frac{1}{2} \sum_{IJ} \frac{\mathrm{d}^2 E}{\mathrm{d}\mathbf{R}_I \mathrm{d}\mathbf{R}_J}\Delta\mathbf{R}_I \Delta\mathbf{R}_J,
\end{equation*}
where $E_0$ is a constant. Demanding that the sum of the first- and second-order terms in the energy are zero at the minimum, gives
\begin{equation*}
\left[ \frac{1}{2}\Delta\mathbf{R}^{\intercal} \mathbf{M} + \nabla_{\mathbf{R}} E \right] \cdot \Delta\mathbf{R} = 0,
\end{equation*}
which, excluding the trivial solution, implies
\begin{equation*}
\Delta\mathbf{R}_{\text{e}}=-2\mathbf{M}^{-1}\nabla_{\mathbf{R}} E,
\end{equation*}
where $\mathbf{M}$ is the matrix of force constants, and $\nabla_{\mathbf{R}} E$ is the multi-atom energy gradient with respect to the configuration atomic displacement vector, $\mathbf{R}$. This is the bare estimate for the atomic displacement correction, up to second order in the energy.
 
\subsection{Correction for global translations}

In order to ensure that the origin of our configuration is fixed and that we have no global translational mode,  we explicitly subtract the center of mass motion of the configuration. 

Given $N_\text{n}$ atoms, each with mass $m_I$, this implies that the global translation correction term is
\begin{equation*}
\Delta\mathbf{R}_{\text{t}}=- \frac{\sum_I m_I \mathbf{R}_I}{\sum_I m_I}.
\end{equation*}
This term is particularly important for non-symmetric molecules, such as hydrogen chloride in \secref{subsec:HCl}.

\subsection{Correction for global rotations}

Similarly, to ensure that the bond length corrections do not result in a rotation of the configuration, or atomic pair rotations, we explicitly subtract global rotational modes. 

The law of moments states that the total moment about the center of mass of any atomic pair, as well as the total moment about the origin of the configuration, is zero, which gives $\sum_I m_I \mathbf{b}_I=\mathbf{0}$ and $\sum_I m_I \mathbf{R}_I=\mathbf{0}$, where $\mathbf{b}$ is the half-bond length between an atomic pair. Together, these relations imply
\begin{equation}
\label{eq:rot1}
\sum_I m_I \mathbf{R}_I \times (\mathbf{b}_I-\mathbf{R}_I\times\boldsymbol{\theta})=\boldsymbol{0},
\end{equation}
where $\mathbf{b}_{\mathrm{r},I}\equiv \mathbf{b}_I-\mathbf{R}_I\times \boldsymbol{\theta}$ is the corrected half bond length to be found. Hence, an expression for the angular displacement of the molecule $\boldsymbol{\theta}_I\equiv(\theta_{\mathrm{x}I},\theta_{\mathrm{y}I},\theta_{\mathrm{z}I})$ is needed. Using the vector triple product identity, we find that Eq.~\ref{eq:rot1} reduces to
\begin{equation*}
\sum_I m_I \mathbf{R}_I \times \mathbf{b}_I = \sum_I \left[ m_I \mathbf{R}_I (\mathbf{R}_I\cdot \boldsymbol{\theta}) - m_I\boldsymbol{\theta}(\mathbf{R}_I \cdot \mathbf{R}_I) \right],
\end{equation*}
which after rearrangement becomes
\begin{widetext}
\begin{equation*}
\underbrace{\sum_I m_I
\begin{pmatrix} R_{\mathrm{y}I}b_{\mathrm{z}I}-R_{\mathrm{z}I}b_{\mathrm{y}I} \\
R_{\mathrm{z}I}b_{\mathrm{x}I}-R_{\mathrm{x}I}b_{\mathrm{z}I} \\
R_{\mathrm{x}I}b_{\mathrm{y}I}-R_{\mathrm{y}I}b_{\mathrm{x}I}
\end{pmatrix}}_{\mathbf{a}} =\underbrace{\sum_I m_I
\begin{pmatrix} -R_{\mathrm{y}I}^2-R_{\mathrm{z}I}^2 & R_{\mathrm{y}I}R_{\mathrm{x}I} & R_{\mathrm{z}I}R_{\mathrm{x}I} \\
R_{\mathrm{x}I}R_{\mathrm{y}I} & -R_{\mathrm{x}I}^2-R_{\mathrm{z}I}^2 & R_{\mathrm{z}I}R_{\mathrm{y}I} \\
R_{\mathrm{x}I}R_{\mathrm{z}I} & R_{\mathrm{y}I}R_{\mathrm{z}I} & -R_{\mathrm{x}I}^2-R_{\mathrm{y}I}^2
\end{pmatrix}}_{\mathbf{B}}
\begin{pmatrix}
\theta_\mathrm{x} \\
\theta_\mathrm{y} \\
\theta_\mathrm{z}
\end{pmatrix}.
\end{equation*}
\end{widetext}
This implies that the atomic displacement correction for global rotations, is
\begin{equation*}
\Delta\mathbf{R}_{\text{r},I}=-\mathbf{R}_I\times\boldsymbol{\theta},
\end{equation*}
where $\boldsymbol{\theta}=\mathbf{B}^{-1}\mathbf{a}$.

\subsection{Correction for anharmonicity}
\label{subsec:corr_anharm}

Up to this point in the analysis, we have assumed that the interaction between atomic pairs is harmonic. Although this is a valid approximation at short distances, at larger distances this approximation breaks down and so a correction term is necessary. One of the most well-studied models used to capture anharmonicity in the interaction between diatomic molecules is the Morse Hamiltonian, which we use as an approximation for our case studies. The Morse Hamiltonian is given by
\begin{equation*}
\hat{H}=\frac{\hat{p}^2}{2\mu}+\hat{V}
\end{equation*}
with a Morse potential
\begin{equation}
\label{eq:Morse_pot}
\hat{V}=V(x)=D[1-e^{-\alpha x}]^2,
\end{equation}
where $D$ is the $x=x_0$ energy minimum depth relative to the dissociation limit at $x\to\infty$ and $\alpha$ determines the curvature of the potential~\cite{Morse29}.

The eigenvalues of the Morse Hamiltonian are
\begin{equation*}
E_n=\hbar\omega_0\left[ \left( n+\frac{1}{2} \right)-x_\text{e} \left( n+\frac{1}{2} \right)^2 \right],
\end{equation*}
where $\omega_0=\sqrt{2D\alpha^2/\mu}$ is the fundamental frequency, $x_\text{e}=\hbar\omega_0/4D$ is the anharmonic constant, and $n\in\mathbb{Z}^+$ is the principal quantum number.
 
Note that the minima of the harmonic and Morse potentials are the same. However, due to the dissociative limit of the Morse potential, the expectation value of position is shifted in the positive x direction in the Morse case. One of the main advantages of this model is that the majority of its properties can be expressed analytically.

By setting $D=\frac{\hbar^2\alpha^2}{2\mu}(N+1/2)^2$, the Morse Hamiltonian may be written as
\begin{equation*}
\hat{H}=-\frac{\hbar^2}{2\mu}\frac{\partial^2}{\partial x^2}+\frac{\hbar^2\alpha^2}{2\mu}\left( N+\frac{1}{2} \right)^2 (e^{-2x}-2e^{-x})
\end{equation*}
up to a constant term. The expectation value of position with respect to the ground state Morse wave function is then
\begin{equation*}
\braket{0|\hat{x}|0}=\frac{\ln(2N+1)-\psi(2N)}{\alpha},
\end{equation*}
where $\psi$ is the digamma function~\cite{Lima05}. Expanding the expectation value of position gives
\begin{equation*}
\braket{0|\hat{x}|0}=\frac{3}{2}\sqrt{\frac{\hbar x_\text{e}}{2\mu \omega_0}}
\end{equation*}
up to leading order in $x_\text{e}$. This is the shift in the equilibrium bond length due to the anharmonicity of the Morse potential.

In order to evaluate this shift, an estimate for the anharmonic constant is needed. Expanding the Morse potential (Eq.~\ref{eq:Morse_pot}) about the equilibrium displacement $x=x_\text{0}$ in powers of $x$, we find that
\begin{equation*}
V(x)=\frac{1}{2}\mu \omega_0^2 x^2+ \sqrt{\frac{\mu^3 x_\text{e} \omega_0^5 }{2\hbar}}x^3+\dots
\end{equation*}
up to a constant term. Comparing quadratic and cubic terms in $x$ with the general form of the Taylor expansion, and solving simultaneously, yields
\begin{equation*}
x_\text{e}=\frac{\hbar}{18\sqrt{\mu}} \left( \left. \frac{\diffd^3 V}{\diffd x^3}\right|_{x_0} \right)^2 \left( \left. \frac{\diffd^2 V}{\diffd x^2} \right|_{x_0} \right)^{-5/2}.
\end{equation*}
Conventionally, the third derivative of the energy is extracted from the curvature of the force, however now utilizing the new information available, we extract the anharmonic constant directly from the gradient of the force constant.

\section{Vibrational modes calculation}
\label{sec:details_vm}

In this section, we describe in detail the methods used to determine the vibrational modes and frequencies of atomic configurations, as well as their associated statistical uncertainties.

\subsection{Exisiting computational approaches}


In order to calculate an estimate for the frequency using the RHF and DFT methods, we use the default scheme, PBE and B3LYP exchange-correlation functionals, respectively, within the CRYSTAL program~\cite{CRYSTAL14}. 

\subsection{Matrix of force constants approach}


The direct method to obtain the vibrational frequencies of a molecule is from an exact diagonalization of the matrix of force constants. Consider, for example, a diatomic molecule in one dimension, such as the hydrogen molecule discussed in \secref{subsec:H}. The matrix of force constants for this system may be written as
\begin{equation}
\label{eq:matrix_cd}
\mathbf{M}=
\begin{pmatrix}
\frac{\mathrm{d}^2E}{\mathrm{d}R_1^2} & \frac{\mathrm{d}^2E}{\mathrm{d}R_1 \mathrm{d}R_2} \\
\frac{\mathrm{d}^2E}{\mathrm{d}R_2 \mathrm{d}R_1} & \frac{\mathrm{d}^2E}{\mathrm{d}R_2^2}
\end{pmatrix}.
\end{equation}
By exactly diagonalizing the matrix, we obtain the eigenmodes, and eigenfrequencies of the system given by
\begin{multline}
\label{eq:omega1}
\omega^2=\frac{1}{2}\left(\frac{\mathrm{d}^2E}{\mathrm{d}R_1^2} + \frac{\mathrm{d}^2E}{\mathrm{d}R_2^2} \right) \\
\pm \sqrt{\frac{1}{4} \left( \frac{\mathrm{d}^2E}{\mathrm{d}R_1^2} - \frac{\mathrm{d}^2E}{\mathrm{d}R_2^2} \right) + \left( \frac{\mathrm{d}^2E}{\mathrm{d}R_1 \mathrm{d}R_2} \right)^2 },
\end{multline}
where the positive frequencies are physical. The errors are calculated using Monte Carlo, as discussed in \secref{subsec:MC_errors}.

There are two possible disadvantages of this method for obtaining the vibrational frequencies of a configuration. First, since it is a complete diagonalization method, it uses all of the entries in the matrix of force constants. However, many of these entries are related by symmetries, and so these calculations are potentially redundant. Second, due to numerical inaccuracy, Eq.~\ref{eq:omega1} may result in an overestimate of the frequencies if the diagonal terms in Eq.~\ref{eq:matrix_cd} are not equal.

Following from the previous example, by imposing the known modes of a diatomic molecule in one dimension, we may write the matrix of force constants as
\begin{equation*}
\mathbf{M}_\text{KM}=\frac{1}{2}
\begin{pmatrix}
\frac{\mathrm{d}^2E}{\mathrm{d}(R_1+R_2)} & 0 \\
0 & \frac{\mathrm{d}^2E}{\mathrm{d}(R_1-R_2)}
\end{pmatrix},
\end{equation*}
which now yields the eigenfrequencies
\begin{equation*}
\omega_\text{KM}^2=\frac{1}{2}\left( \frac{\mathrm{d}^2E}{\mathrm{d}R_1^2} + \frac{\mathrm{d}^2E}{\mathrm{d}R_2^2} \right)
\pm \left( \frac{\mathrm{d}^2E}{\mathrm{d}R_1 \mathrm{d}R_2} \right).
\end{equation*}
Notice that $|\omega_\text{KM}|\leq|\omega|$ due to the absence of the diagonal terms in the square root of Eq.~\ref{eq:omega1}. 

For a general system, we may input a set of known modes $\{\mathbf{x} \}$. These $3N_\text{n}$-dimensional row vectors act on the $3N_\text{n}\times 3N_\text{n}$ dynamical matrix, $\mathbf{D}$, to extract the corresponding eigenfrequency, such that
\begin{equation*}
\omega_{\text{KM},i}=\hat{\mathbf{x}}_i\mathbf{D}\hat{\mathbf{x}}_i^{\intercal},
\end{equation*}
with corresponding error
\begin{equation*}
\sigma_{\text{KM},i}=\sqrt{\hat{x}_{ij} (2\Sigma^2_{jk} - \Sigma_{jk} \delta_{jk})\hat{x}_i^k },
\end{equation*}
where the hats denote normalization, $\boldsymbol{\Sigma}$ is the standard error matrix corresponding to $\mathbf{M}$, and the dynamical matrix, $\mathbf{D}$, is the matrix of force constants weighted by the atomic masses.

By imposing known modes on the system, we can reduce the potential for numerical error and speed up the matrix diagonalization. However, these advantages only hold if the correct eigenmodes are known a priori, and therefore we do not employ this scheme as standard for our DMC calculations.

\subsection{Approaches based on derivatives of the force and energy}
%

Further to the methods based on the matrix of force constants, we also consider traditional techniques, for comparison. 

We obtain an estimate of the frequency ($\omega_\text{FG}$) from the gradient, $\kappa$, of the interatomic force against bond length graph. The error in the gradient of the slope is the asymptotic standard error from a linear regression, and this is propagated to the vibrational frequency in the usual way:
\begin{equation*}
\sigma_{\omega}^2=\left| \frac{\partial \omega}{\partial \kappa} \right|^2 \sigma_\kappa^2.
\end{equation*}

Similarly, an additional estimate of the vibrational frequency ($\omega_\text{EC}$) is obtained by computing the second derivative of the energy at a series of displacements along the trajectory of an eigenmode. For this, we use a numerical central difference scheme. Since this result is based on a linear superposition of energy data points, the errors add in quadrature.

\subsection{Correction for anharmonicity}

All of the above methods for calculating the vibrational frequency rely on the harmonic potential approximation. However, there are certain cases where anharmonic vibration is dominant and a correction to these frequencies needs to be applied. As for atomic relaxation, we apply an approximate correction, due to a Morse potential, which for the fundamental vibrational frequency, is given as $\Delta\omega=-x_\text{e}/4$, where $x_\text{e}$ is the anharmonic constant.

\subsection{Monte Carlo uncertainty}
\label{subsec:MC_errors}

The matrix of force constants $\mathbf{M}$ comes with an associated standard error matrix, $\boldsymbol{\Sigma}$, from the reblocking method in CASINO~\cite{Flyvbjerg89}. Calculating the errors in eigenvalues given the errors in the matrix elements is a non-trivial task and one which has been studied extensively in pure mathematics~\cite{Alefeld86,Henning88,Dongarra83,Rump89,Yamamoto80,Yamamoto82}. For the purposes of this paper, we calculate the eigenvalue errors using Monte Carlo.

For each Monte Carlo run we generate a dynamical matrix, whose matrix elements are normally distributed, with a mean equal to the original matrix elements and a standard deviation equal to the corresponding standard errors. We then perform many runs until the average eigenvalues converge to the true eigenvalues, and we use the standard errors of these Monte Carlo runs as the errors in the eigenvalues.


\bibliographystyle{apsrev4-1}
\bibliography{mfc}

%
%
%

\end{document}